# A thermodynamic parallel of the Braess road-network paradox

Kamal Bhattacharyya*
Department of Chemistry
92, A. P. C. Road, Kolkata 700 009, India

**Abstract**

We provide here a thermodynamic analog of the Braess road-network paradox with irreversible engines working between reservoirs that are placed at vertices of the network. Paradoxes of different kinds reappear, emphasizing the specialty of the network.

## 1. Introduction

The Braess paradox (BP) [1] for a road network, to state in terms of a very simple case, concerns the inflow of cars from point 1 and outflow through point 4 (see Figure 1).

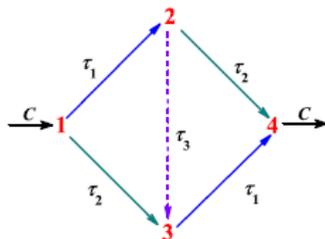

**Figure 1. The basic network where path 23 is the extra path.**

The pair of roads 12 and 34 has similar characteristics, so is true of the other pair 24 and 13. Cars (*C*) can go through roads 124 and 134. When points 2 and 3 are joined by another possible path, to be henceforth called the *extra path* (EP), cars have more freedom. But, it has been shown that this freedom may lead to *increased cost* in certain situations, contrary to common sense. Note that the conventional belief is, *freedom is always an advantage*. This maxim is shattered here, rather surprisingly. And, that's why, it is paradoxical. In summary, *addition of an extra route within a network may not be profitable*.

With reference to the above figure, let us analyze the various possibilities to check when and how such a paradox might emerge. To proceed, we need to define a characteristic time ($\tau$) for each path. There exist several options. We stick to one popular set, where $\tau_1 = C/100$, $\tau_2 = 45$ and $\tau_3 = 1$, in some unit of time (say, minute). Such a choice implies, roads 12 and 34 are substantially narrower. Thus, jamming would cost a longer time as more cars pass through them. The other pair of routes is sufficiently wider, but longer too, to allow (within limits) the passage of any number of cars at fixed times. Path 23 (the EP), on the other hand, is widest and shortest. The standard choice is to take 4000 total cars ($C_T = 4000$) to illustrate the problem. There are two possibilities. A driver may exercise his own choice of route, with or without any knowledge of what others are doing; we call it the user-optimized (UO) travel. Otherwise, the choice is dictated by the traffic administration, keeping explicit records (with automatic display) of the number of cars entering and leaving a particular path so as to minimize the travel time, and this is termed as the system-optimized (SO) scheme. To succinctly present the paradox, we now consider Table 1 where results of sticking to several alternative routes are displayed.

*kbchem@caluniv.ac.in; pchemkb@gmail.com

Our starting premise is again the popular choice of $C_T = 4000$, but we also vary the number to notice certain other effects. A glance at Table 1 reveals the following:

**Table 1. Illustration of the Braess paradox. Here $C_T$ and $\tau_T$ refer respectively to the total number of cars and the total time required. $C_{ij}$ defines the number of cars along a path $ij$. System-optimized (SO) cases are distinguished from user-optimized (UO) ones. Characteristic times ($\tau$) for the paths are defined in the text. Nash equilibrium (NE) prevails in all the cases.**

| $C_T$ | $C_{12}$ | $C_{13}$ | $C_{23}$ | $C_{24}$ | $C_{34}$ | $C_{123}$ | $C_{124}$ | $C_{134}$ | $C_{1234}$ | $\tau_T$ | Remark |
|---|---|---|---|---|---|---|---|---|---|---|---|
| 4000 | 2000 | 2000 | 0 | 2000 | 2000 | 0 | 2000 | 2000 | 0 | 65 | SO, NE |
| 4000 | 4000 | 0 | 4000 | 0 | 4000 | 4000 | 0 | 0 | 4000 | 81 | UO, NE |
| 4000 | 2200 | 1800 | 2200 | 0 | 4000 | 2200 | 0 | 1800 | 2200 | 63 | SO, NE |
| 2000 | 1000 | 1000 | 0 | 1000 | 1000 | 0 | 1000 | 1000 | 0 | 55 | SO, NE |
| 2000 | 2000 | 0 | 2000 | 0 | 2000 | 2000 | 0 | 0 | 2000 | 41 | SO, UO, NE |
| 6000 | 3000 | 3000 | 0 | 3000 | 3000 | 0 | 3000 | 3000 | 0 | 75 | SO, NE |
| 6000 | 2200 | 3800 | 2200 | 0 | 6000 | 2200 | 0 | 3800 | 2200 | 83 | UO, NE |
| 4400 | 2200 | 2200 | 0 | 2200 | 2200 | 0 | 2200 | 2200 | 0 | 67 | SO, NE |
| 4400 | 2200 | 2200 | 2200 | 0 | 4400 | 2200 | 0 | 2200 | 2200 | 67 | SO, UO, NE |

(i) When the EP 23 is absent, the first row shows how an intelligent system can optimize the 'cost' $\tau_T$. Any other choice would only increase the overall cost for *all* the cars.

(ii) If the EP is operative, the drivers might follow a simpler logic of *lessening the cost* by choosing the route 123 instead of the direct path 13. Results are shown in the second row, and one may wonder that this leads finally to a *loss* (compare against the first row). Here we witness the impact of the paradox.

(iii) The system may still try to optimize, and it can do so by following the third row of Table 1, *i.e.*, allowing 2200 cars to go through path 1234 in a time $\tau \leq \tau_2$. This provides the 1800 cars at point 3 (after $\tau_2$) a decisive advantage of *far less* crowding as they cross the path 34. Such a tactic is *never emphasized*, but it is surely admissible.

(iv) The EP may always be profitable with *lesser* total cars. The fourth and fifth rows testify this fact. Further, the UO and SO become equivalent in this case.

(v) However, with *much larger* $C_T$, even the new trick of the SO scheme (*cf.* the third row) would cease to work. This is clear from the results displayed in the last two rows. Here, the BP is operative in full glory. Thus, along with $\tau$ for the different paths, $C_T$ plays a crucial role.

(vi) Rows 8 and 9 reveal the limiting value of $C_T$ *beyond* which BP shows up.

The above table also highlights the importance of *flux*. Note that, of all the cases under consideration here, the flux is *maximum* at the *sixth* row and, therefore, the BP is very relevant to transportation science in complex road networks. Particularly if the flux is large, EP becomes uneconomic. Further, *if the system does not properly manage, the EP may cause serious traffic jams*. The added flexibility (EP 23) thus acts adversely. It is in such a situation that its construction becomes questionable [2]! Let us note in passing that, *only for small $C_T$* [like 2200 or less], the route 1234 is rewarding, and the more so in terms of flux as $C_T$ increases from its minimum possible value of unity.

A related concept in the above issue is the Nash equilibrium (NE) [3]. It says that, when such an equilibrium exists, one cannot gain in cost (*i.e.*, lessen the travel time) by adopting a different route, knowing others' strategies. As noted in Table 1, the NE is always valid in all the suggested moves with well-defined number of cars ($C_T$).

The BP stimulated a lot of further works (see, e.g., [4-8] and references quoted therein). While road networks had been the prime concern [4], other types of networks [5], including chemical ones [6], have also shown its importance. Transport in nanostructures [7] and epidemiology [8] also highlighted some relevance of the BP.

Having understood the nature of the paradox, let us emphasize now that the network depicted in Figure 1 is itself a *very special* one. Thus, paradoxes akin to the BP may arise in other areas too. As a test case, we expound here a thermodynamic parallel.

**2. A thermodynamic network**

To proceed, we consider 4 reservoirs, to be designated as 1, 2, 3 and 4, kept at temperatures $T_1 > T_2 > T_3 > T_4$. Let us choose also 4 *irreversible* engines ($\eta_k < \eta_{CE}$), each working between 2 such reservoirs with a given efficiency, as shown in Figure 2. Efficiency $\eta_{CE}$ refers to the Carnot engine (CE) acting between the corresponding reservoirs. The symmetry in efficiencies is deliberate, so also the arrangement (see, *e.g.* Figure 1). The source is at the left ($T_1$) and the sink is at the right ($T_4$). The other two reservoirs act merely as intermediate-temperature ones that finally do not gain or lose any heat. Thus, they do not contribute to the overall entropy change.

The idea has its origin in the following argument. Suppose we have two reservoirs at $T_1$ and $T_4$, and an engine operates between them with a fixed efficiency $\eta_1$, much less than the Carnot efficiency. Then, the work obtainable is also $\eta_1$ per *unit* amount of heat taken from the source. Had it been replaced by another engine of efficiency $\eta_2$, work would be just $\eta_2$. But, if we insert a third intermediate-temperature reservoir ($T_2$), which does not take up or lose any heat in a cycle, and allow both the engines to operate (path 124), satisfying $\eta_k < \eta_{CE}$, but keeping the efficiencies same as before, work $W$ (see below) becomes greater. Within limits of $\eta_k < \eta_{CE}$, the temperature of this third reservoir does not influence $W$. So, we add another such reservoir at $T_3$ (path 134) and allow alterations of the two engines to check the outcomes. As we now have two intermediate reservoirs at different temperatures, it is also tempting to connect them by another engine. The whole arrangement thus follows. The physical connotation is different here, but the resemblance with Figure 1 is immediate. Anyway, let us now proceed to the *cost* estimates. Cost may be defined here as the inverse of work, since we like to extract *maximum work*; the cost is thus minimized.

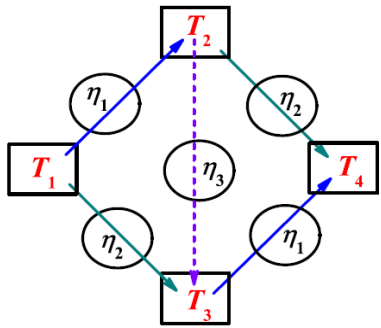

**Figure 2. A network of reservoirs connected by irreversible engines.**

There is one more difference. Whereas in the original road network, we varied the number of cars (equivalently, the flux), here we vary the efficiencies of the engines. The chief aim remains same, however, *viz.*, the *cost-effectiveness*.

Assume, as stated earlier, that the heat taken from the source is *unity*. One may check that the following results for work $W$ then emerge (paths are denoted by temperature subscripts):

A: Path 124
$$W_{124} = \eta_1 + (1-\eta_1)\eta_2 = W. \qquad (1)$$

B: Path 134
$$W_{134} = \eta_2 + (1-\eta_2)\eta_1 = W. \qquad (2)$$

C: Path 124 + Path 134
$$\begin{aligned} W_{124+134} &= x_1\eta_1 + x_1(1-\eta_1)\eta_2 + (1-x_1)\eta_2 + (1-x_1)(1-\eta_2)\eta_1 \\ &= \eta_1 + \eta_2 - \eta_1\eta_2 = W. \end{aligned} \qquad (3)$$

In (3), by our choice, $x_1$ fraction of heat flows through path 124, but $x_1$ does not have any effect on the net work finally. Thus, cases A, B and C yield the *same* work. In other words, there is *no gain* in terms of work. As a result, we have the first paradox, very similar to the BP in spirit:

*Paradox* 1: *Addition of a freedom may not always be profitable.*

In the above statement, we mean by 'freedom' the path 134 that is added here to the existing path 124. This forms the simplest network. But, it has not turned out to be useful because the effective efficiencies along both the paths are same, and the total heat is apportioned between the routes. Hence, one of the intermediate-temperature reservoirs seems now *redundant*. However, we are now close to another more serious paradox.

*Paradox* 2: *How does one know that all the engines are functional?*

This problem arises because, it is not known for sure that *all* cases A, B and C are *active simultaneously*. For example, it may well occur that the engines along paths 13 and 34 *remain idle* throughout (*i.e.*, $x_1 = 1$, by choice). There is no theoretical principle to refute this possibility. Similarly, $x_1 = 0$ would mean that the other two engines choose to stay *inactive*. We notice that our stress on the *specialty* of the network in Figure 1 has now found some significance once one appreciates this second paradox.

To continue further, we insert an extra engine between $T_2$ and $T_3$. In presence of this extra path (EP), we have one more possibility, as summarized below:

D: Path 1234

$$\begin{aligned} W_{1234} &= \eta_1 + (1-\eta_1)\eta_3 + (1-\eta_1)(1-\eta_3)\eta_1 \\ &= W + (1-\eta_1)[\eta_1 + \eta_3(1-\eta_1) - \eta_2] \\ &= W + (1-\eta_1)\Delta\eta. \end{aligned} \qquad (4)$$

Here, the term within the bracket at the right side of the second line is designated by $\Delta\eta$. It is the difference in work of paths 123 and 13. This difference in cost was central to the original paradox as well [see point (ii) below Table 1]. Result (4) shows that (*a*) $\Delta\eta = 0$ refers to something similar to an equilibrium situation where cases A, B, C and D act equivalently, (*b*) $\Delta\eta > 0$ sets a preference of case D over the other three in respect of available work, (*c*) case D involves a larger entropy change for $\Delta\eta < 0$, rendering less work, and (*d*) a *sufficient* condition for $\Delta\eta > 0$ is to ensure $\eta_2 \leq \eta_1$, while $\eta_2 > \eta_1$ is a *necessary* condition for $\Delta\eta < 0$.

One may ascertain easily that all the above situations are permissible. Indeed, Table 2 summarizes such cases for specific choices of the temperatures of reservoirs and efficiencies of engines attached, as per the construction of Figure 3. Such efficiencies are all less than the corresponding Carnot efficiencies to guarantee their irreversible characters. Note that the sample values of $\eta_k$ in Table 2 are purely arbitrary; there exists countless of possibilities (within limits, of course) yielding positive, negative or zero $\Delta\eta$.

**Table 2. Typical cases of positive, negative and zero $\Delta\eta$.**

| $T_1$ | $T_2$ | $T_3$ | $T_4$ | $\eta_1$ | $\eta_2$ | $\eta_3$ | $\Delta\eta$ |
|---|---|---|---|---|---|---|---|
| 400 | 200 | 100 | 50 | 1/4 | 1/6 | 1/3 | 1/3 |
| 400 | 200 | 100 | 50 | 1/10 | 1/3 | 1/6 | -1/12 |
| 400 | 200 | 100 | 50 | 1/10 | 1/4 | 1/6 | 0 |

For example, at fixed $\eta_3 = 4/9$, these requirements may be fulfilled over a wide range of $\eta_2$ and $\eta_1$, as displayed in Figure 4. The guiding relation in this regard is (4). We also see that points on the boundary correspond to $\Delta\eta = 0$.

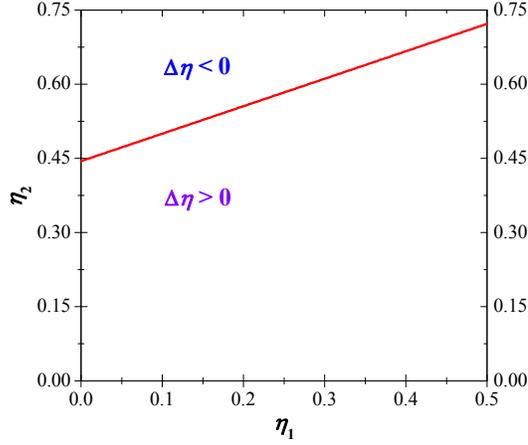

**Figure 3. Boundary separating $\Delta\eta > 0$ and $\Delta\eta < 0$ at $\eta_3 = 4/9$.**

Assume further now that *all paths* are allowed, including the EP. Suppose $x_2$ fraction of heat gained by reservoir 2 goes along the 24 path. We then have the following result after some simplification:

$$W_{EP} = W + x_1(1-x_2)(1-\eta_1)[\eta_1 + \eta_3(1-\eta_1) - \eta_2]. \tag{5}$$

It shows that the work is now *different* in presence of the EP. Rewriting (5) as

$$W_{EP} = W + x_1(1-x_2)(1-\eta_1)\Delta\eta, \tag{6}$$

one finds again three distinct situations with positive, negative and zero $\Delta\eta$.

Collecting all the above results, with the restrictions about the efficiencies, the arrangements and role of the intermediate reservoirs, we now come to certain additional interesting conclusions.

Let us start with the case of $\Delta\eta = 0$. This is going to be certainly fruitless because here one *does not gain at all*:

$$W_{EP} = W = W_{1234}. \tag{7}$$

On the other hand, $\Delta\eta > 0$ case seems to be *most favorable*, yielding from (4) and (6)

$$W_{1234} \geq W_{EP} \geq W. \tag{8}$$

Similarly, when $\Delta\eta < 0$, one is led to

$$W_{1234} \leq W_{EP} \leq W. \tag{9}$$

Indeed, the equality $W_{EP} = W_{1234}$ is achieved only when $x_1 = 1$ *and* $x_2 = 0$, which means that the *two routes become identical*. Note, however, that (6) generally contains two unknown variables $x_1$ and $x_2$ for any $\Delta\eta \neq 0$. In order to get rid of these variables, one may choose to put in (6) the 'average estimates' of $x_1$ and $x_2$. If we take an assembly of the set-up as in Figure 2, such an averaging becomes meaningful because both the variables are allowed to take random values between 0 and 1. We thus have

$$\bar{W}_{EP} = W + \tfrac{1}{4}(1-\eta_1)\Delta\eta. \tag{10}$$

Some more paradoxical situations are now in order:

*Paradox* 3: *When* $\Delta\eta = 0$, *how does one know that all cases* A, B, C *and* D *are active, and hence all the engines are again functional*?

This is similar to the second paradox, but it now *includes* path 1234. To appreciate the extension, notice that the inclusion of $x_2$, and hence the possibility of its being zero or unity, *adds to confusion*. Again, one wonders if there exists any guiding principle in this regard.

*Paradox* 4: *If* $\Delta\eta > 0$, *is it guaranteed that case* D *will be followed to offer maximum work*?

One might have expected this answer in the affirmative. But, there are two immediate objections in view of (8) and (10). Additionally, the problem may be *even more deceptive*. We know that *a process with larger increase in entropy is more probable*. So, if flexibility exists, nature will try to gain the maximum profit out of it by *maximizing* the entropy of the *surroundings*. Therefore, $W_{EP}$ will indeed try to be a *minimum*. (As $W$ does not have any freedom, it remains a constant, anyway.) In other words, the $\Delta\eta > 0$ case will avoid the 23 path fully ($x_1 = 0$ or $x_2 = 1$) to meet the equality. On the other hand, for $\Delta\eta < 0$, the wisest choice would be $x_1 = 1$ and $x_2 = 0$ so as to *depart most* from the equality. Here, the EP incurs a real loss! This is closest in spirit to the Braess' paradox. In a thermodynamic context, we thus see that, in place of the system, the surroundings would take advantage of the extra freedom. So, when $\Delta\eta > 0$ and we like to exploit the profit, *the network needs to be broken* in the way shown in Figure 4. This ensures only case D to act.

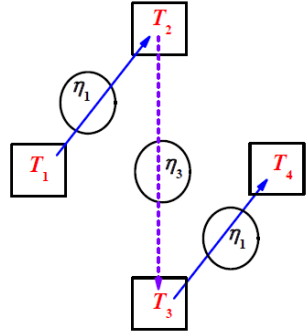

**Figure 4. Part of the network for profitable use when Δ$\eta$ > 0.**

**3. Conclusion**

In summary, the analogy with a flow system, as prevalent in a road network, follows as soon as we define that *unit* amount of heat is released from the source *per unit time* in our given arrangement. Then, of course, the work will be replaced by *power*, but all our above statements will be valid. Finally, Δ$\eta$ = 0 if all the engines are CE, and hence no paradox would arise.

**Acknowledgement**

I am grateful to Dr. S. N. Bhattacharyya for his kind interest and a careful reading of the article.